\documentclass[a4paper,fleqn,usenatbib]{mnras}

\usepackage{newtxtext,newtxmath}
\usepackage[T1]{fontenc}
\usepackage{ae,aecompl}
\usepackage{graphicx}	
\usepackage{amsmath}	
\usepackage{amssymb}	
\usepackage{enumitem}
\usepackage{rotating}

\newcommand{\gc}{$\gamma$\,Cas}
\newcommand{\xmm}{{\sc{XMM}}\emph{-Newton}}

\title[3 new \gc\ stars]{Three discoveries of \gc\ analogs from dedicated \xmm\ observations of Be stars\thanks{Based on data obtained with \xmm , an ESA Science Mission with instruments and contributions directly funded by ESA Member States and the USA (NASA). Also based on spectra obtained with the TIGRE telescope, located at La Luz observatory, Mexico (TIGRE is a collaboration of the Hamburger Sternwarte, the Universities of Hamburg, Guanajuato, and Li\`ege).}}

\author[Y. Naz\'e et al.]{Ya\"el~Naz\'e$^1$\thanks{F.R.S.-FNRS Senior Research Associate, email: ynaze@uliege.be}, Christian Motch$^{2}$, Gregor Rauw$^1$, Shami Kumar$^1$, Jan Robrade$^{3}$, \newauthor{Raimundo Lopes de Oliveira$^{4,5}$, Myron A. Smith$^{6}$, and Jos\'e M. Torrej\'{o}n$^{7}$}
\\
$^{1}$Groupe d'Astrophysique des Hautes Energies, STAR, Universit\'e de Li\`ege, Quartier Agora (B5c, Institut d'Astrophysique et de G\'eophysique), \\
All\'ee du 6 Ao\^ut 19c, B-4000 Sart Tilman, Li\`ege, Belgium\\
$^{2}$Universit\'e de Strasbourg, CNRS, Observatoire Astronomique de Strasbourg, 11 rue de l'Universit\'e, F-67000 Strasbourg, France\\
$^{3}$Hamburger Sternwarte, University of Hamburg, Gojenbergsweg 112, 21029 Hamburg, Germany\\
$^{4}$Departamento de F\'isica, Universidade Federal de Sergipe, Av. Marechal Rondon, S/N, 49000-000 S\~ao Crist\'ov\~ao, SE, Brazil\\
$^{5}$Observat\'orio Nacional, Rua Gal. Jos\'e Cristino 77, 20921-400, Rio de Janeiro, RJ, Brazil\\
$^{6}$NSF OIR Lab, 950 N Cherry Ave, Tucson, AZ 85721, USA\\
$^{7}$Instituto Universitario de F\'{\i}sica Aplicada a las Ciencias y las Tecnlog\'{\i}as, Universidad de Alicante, E-03690 Alicante, Spain
}

\pubyear{2019}

\begin{document}
\label{firstpage}
\pagerange{\pageref{firstpage}--\pageref{lastpage}}
\maketitle

\begin{abstract}
In the last years, a peculiarity of some Be stars - their association with unusually hard and intense X-ray emission - was shown to extend beyond a mere few cases. In this paper, we continue our search for new cases by performing a limited survey of 18 Be stars using \xmm. The targets were selected either on the basis of a previous X-ray detection ({\it Exosat, ROSAT, XMM}-slew survey) without spectral information available, or because of the presence of a peculiar spectral variability. Only two targets remain undetected in the new observations and three other stars only display faint and soft X-rays. Short-term and/or long-term variations were found in one third of the sample. The spectral characterization of the X-ray brightest 13 stars of the sample led to the discovery of three new \gc\ (HD\,44458, HD\,45995, V558\,Lyr), bringing the total to 25 known cases, and another \gc\ candidate (HD\,120678), bringing the total to 2.
\end{abstract}


\section{Introduction}
Most stars are X-ray emitters, though their X-ray luminosities span a wide range of values. For the cases that display particularly intense emission, the X-rays constitute an important probe of the physical processes occurring in those objects. In massive stars, stellar winds usually play the leading role in generating X-rays. Being intrinsically unstable, these line-driven winds possess shocks distributed through the outflow which generate soft X-rays (typically $kT\sim$0.6\,keV) with an intensity following the ``canonical'' $\log (L_X/L_{BOL})=-7$ relation in O-stars and very early B-stars \citep[e.g.][]{ber97,naz11}. As the effective temperature decreases, winds become weaker and weaker, hence most B-stars appear X-ray faint. There are several exceptions, however, and the most common ones are: a strong dipolar magnetic field may channel the wind flows towards the equator where they collide, generating bright and moderately hard X-rays \citep{bab97,udd14,naz14}; an otherwise unseen PMS companion generates sufficient X-rays, notably during flares, to lead to a detection of the system \citep[e.g.][]{san06}; the presence of an accreting compact companion leads to the emission of very hard and very intense X-rays, in particular in Be X-ray binaries \citep{rei11}. Finally, there is the so-called \gc\ category \citep{smi16}.

\gc\ analogs, named after their prototype, are first of all Be stars, i.e. they possess (or have possessed) a circumstellar Keplerian decretion disk whose signature can be seen through strong emission lines in the optical spectrum (for a review on Be stars, see \citealt{riv03}). In the optical range, up to now, they do not seem to particularly stand out amongst the Be family. The defining criteria of these objects come from the X-ray range (for a review, see \citealt{smi16}). At high energies, \gc\ analogs display a thermal spectrum associated to a high plasma temperature ($kT\sim5-20$\,keV, i.e. much hotter than found in ``normal'' and magnetic B-stars). Furthermore, their X-ray luminosities are intermediate between those of ``normal'' massive stars and those of X-ray binaries ($\log[L_{\rm X}/L_{\rm BOL}]\sim -5$, $L_{\rm X}^{ISM\,cor}(0.5-10\,{\rm keV})=4\times 10^{31}-2\times 10^{33}$\,erg\,cm$^{-2}$\,s$^{-1}$). Finally, they also display short, flaring-like variations of their X-ray emission, as well as long-term changes. 

The origin of these peculiarities remains debated. They cannot be due to the presence of strong dipolar magnetic fields, as in confined winds, since the presence of such fields is incompatible with a Keplerian decretion disk, as has been demonstrated both observationally and theoretically \citep{gru12,udd18}. However, whilst large-scale magnetic fields can be ruled out, localized small-scale fields could exist \citep{can11} and may interact with instability-generated fields of the disk, leading to flaring X-ray emission \citep{rob02}. Alternative explanations involve accretion under unusual conditions onto a compact companion (white dwarf, \citealt{mur86,ham16,tsu18} or neutron star with a propeller process, \citealt{pos17}). 

Up to now, the data remain scarce as many \gc\ stars were discovered by chance. The exact incidence rate of such stars is therefore unclear, with poorly-defined limits (though it is significantly higher than for Be X-ray binaries, see section 6.1 of \citealt{smi17}). Moreover, so far it remains unclear what the physical characteristics and stellar properties that make a Be star display a \gc\ behaviour are. Better statistics are thus eagerly needed, which is why specific searches have been undertaken. \citet{neb13,neb15} searched for Be counterparts to unidentified X-ray sources while \citet{naz18} tackled the problem the other way around, searching data archives for serendipitous X-ray observations of Be stars. All these efforts led to the detection of about twenty \gc\ analogs.

In this paper, we continue this endeavour and report on an X-ray survey of selected Be stars. The choice of targets was done based on two criteria. The first one was a previous detection of the star in the X-ray range. This reveals that the star emits X-rays but it is not a precise characterization of the emission at high energy. Without any available spectrum, the determination of the potential \gc\ character cannot be done. In this context, we cross-correlated the Be Star Spectra catalog \citep[BeSS,][]{nei11} and bright ($V<6$\,mag) Be stars listed in Simbad with {\it ROSAT} OB-stars detections \citep{ber96}. We kept sources displaying $\log (L_{\rm X}/L_{\rm BOL})>-7$ and for which no other X-ray observation was available. We added sources detected by {\it Exosat}-ME in the hard X-ray range but not by {\it ROSAT} in the soft band, as was the case of the \gc\ analog HD\,45314 \citep{rau13,rau18}. Such a situation suggests that these objects display very hard X-ray emissions hence could have a good chance of being \gc\ analogs, if the {\it Exosat} detection was not due to optical/UV loading. A last target, $\alpha$\,Ara, was detected as an X-ray source in the \xmm\ slew survey, as had been the case of the \gc\ analog $\pi$\,Aqr \citep{naz17}. The second criterion relies on the presence of an optical peculiarity observed in \gc\ and HD\,45314 \citep{rau18}: a transition of the H$\alpha$ line from a classical double-peaked pure emission profile to a so-called shell profile. For steady Be disks, a shell profile is observed when the (equatorial) disk is seen edge-on with respect to our line-of-sight. Yet, a small subset of Be stars have displayed shell episodes during which the H$\alpha$ line changed from a conventional double-peaked pure emission to a shell morphology \citep[see Sect. 6.1.2 in][and references therein]{rau18}. This suggests a temporarily more complex geometry of the circumstellar envelope. Since several of the few Be stars displaying these spectacular variations actually belong to the \gc\ category, we included in our sample two other stars having undergone an episode of such variations (HD\,120678 and 59\,Cyg) in the past.

In total, 18 targets were thus selected and observed using \xmm. Section 2 presents the observations and their reduction, Section 3 lists the obtained individual results and compares them with previous observations of Be stars, and Section 4 summarizes our findings and concludes this paper.

\section{Observations and data reduction} 
The limited survey of Be stars was performed with \xmm\ in 2018 and 2019 for our programs 082031 and 084020. These observations were taken in various mode and filter\footnote{In a few observation files, the filter was incorrectly set to ``CalThick'' but we corrected files manually, setting it to the actual ``Thick'' filter identification.} combinations, chosen to avoid potential X-ray pile-up and optical/UV loading. The list of the exposures is available in Table \ref{journal}. This table also provides information on the stellar properties, derived as in \citet{naz18}. Spectral types come from the BeSS or Simbad databases, $V$ magnitudes from Simbad, and interstellar color excess $E(B-V)$ from \citet{cap17} considering the distances. Bolometric luminosities were derived considering bolometric corrections of \citet{nie13} and the absolute magnitudes of \citet{weg07}. The distance intervals were taken from \citet{bai18}, except for $\mu$\,Lup for which the {\it GAIA} DR2 parallaxes were used and the optically bright $\eta$\,Ori, $\alpha$\,Ara, and Sheliak ($\beta$\,Lyr) for which the {\it Hipparcos} distances were used as the {\it GAIA} distances remain uncertain \citep{dri19}. 

\begin{table*}
\centering
\caption{Journal of the \xmm\ observations. A ``:'' at the end of the name indicates a detection to be confirmed and the second column indicates the origin of the target's choice (``e,r,s,x'' for {\it Exosat} detection, {\it ROSAT} detection, spectral behaviour, \xmm\ detection, respectively), see text for details. The specified duration corresponds to MOS1 performed duration, before any flare filtering.}
\label{journal}
\scriptsize
\setlength{\tabcolsep}{3.3pt}
\begin{tabular}{lcllccccccccc}
\hline\hline
Name & & ObsID & mid-observation date & Dur. & Sp.T & $V$ & $E(B-V)$ & $d$ & $\log(L_{\rm BOL}/L_{\odot})$ & \multicolumn{3}{c}{Count rates in 0.3--10.\,keV ($10^{-2}$\,ct\,s$^{-1}$)}\\
 & & & & (ks) & & \multicolumn{2}{c}{(mag)} & (pc) & & pn & MOS1 & MOS2\\
\hline
HD\,18552        & r& 0820310901 & 2018-08-29T02:59:24  & 13.7 & B8Vne    & 6.12 & 0.040 & 227$\pm$6  & 2.48$\pm$0.02 & 18.0$\pm$0.5 & 5.25$\pm$0.26 & 4.85$\pm$0.23\\
$\eta$\,Ori      & r& 0840200301 & 2019-09-16T16:01:25  &  9.0 & B0.5Ve   & 3.38 & 0.006 & 333$\pm$107& 4.64$\pm$0.29 & 22.0$\pm$0.7  & 4.04$\pm$0.29 & 4.04$\pm$0.26\\
HR\,1847         & r& 0820311001 & 2019-03-20T12:30:38  & 17.3 & B7IIIe   & 6.06 & 0.154 & 434$\pm$19 & 3.30$\pm$0.04 & 1.70$\pm$0.16 & 0.58$\pm$0.08 & 0.47$\pm$0.07\\
HD\,43285:       & r& 0820310801 & 2019-03-31T00:55:11  & 10.5 & B5IVe    & 6.05 & 0.016 & 223$\pm$4  & 2.78$\pm$0.02 & 6.32$\pm$0.36 & 1.33$\pm$0.16 & 1.97$\pm$0.17\\
HD\,44458        & r& 0820310301 & 2018-09-08T02:14:29  & 14.6 & B1Vpe    & 5.59 & 0.107 & 619$\pm$42 & 4.39$\pm$0.06 & 51.5$\pm$0.8  & 17.5$\pm$0.5  & 16.6$\pm$0.4 \\
HD\,45995        & r& 0820310401 & 2018-09-24T14:07:40  &  5.5 & B2Vnne   & 6.14 & 0.103 & 662$\pm$44 & 4.03$\pm$0.06 & 31.4$\pm$1.0  & 9.27$\pm$0.53 & 10.5$\pm$0.5 \\
I\,Hya:          & r& 0820310701 & 2018-06-10T08:18:44  & 15.1 & B5Ve     & 4.75 & 0.014 & 153$\pm$7  & 2.91$\pm$0.04 & 26.1$\pm$0.6  & 7.31$\pm$0.30 & 6.93$\pm$0.26\\
QY\,Car          & e& 0840200101 & 2019-06-26T07:18:41  & 11.7 & B2IVnpe  & 5.76 & 0.050 & 474$\pm$25 & 3.87$\pm$0.05 & 0.42$\pm$0.12 & 0.01$\pm$0.02 & 0.22$\pm$0.06\\
$\kappa$\,Dra    & e& 0840200201 & 2019-11-19T11:01:21  & 12.6 & B6IIIpe  & 3.88 & 0.012 & 141$\pm$7  & 3.10$\pm$0.04 & \multicolumn{3}{c}{undetected} \\
HD\,120678       & s& 0820310601 & 2019-03-10T09:11:24  & 42.6 & O9.5Ve   & 8.20 & 0.421 &2344$\pm$218& 5.10$\pm$0.08 & 3.01$\pm$0.56 & 1.58$\pm$0.17 & 0.90$\pm$0.26\\
$\mu$\,Lup       & r& 0820311201 & 2019-03-08T14:32:58  & 12.5 & B8Ve     & 4.27 & 0.009 & 104$\pm$8  & 2.49$\pm$0.06 & \multicolumn{3}{c}{undetected} \\
d\,Lup:          & r& 0820310201 & 2018-08-17T16:42:01  &  6.7 & B3IVpe   & 4.54 & 0.010 & 138$\pm$9  & 3.15$\pm$0.06 & 0.75$\pm$0.20 & 0.10$\pm$0.06 & 0.13$\pm$0.07\\
$\alpha$\,Ara    & x& 0820310101 & 2018-10-08T00:28:13  & 13.3 & B2Vne    & 2.95 & 0.006 &  83$\pm$6  & 3.38$\pm$0.06 & 0.75$\pm$0.15 & 0.12$\pm$0.06 & 0.14$\pm$0.05\\
$\alpha$\,Ara    & x& 0820311501 & 2019-03-03T23:30:46  &  8.7 & B2Vne    & 2.95 & 0.006 &  83$\pm$6  & 3.38$\pm$0.06 & 1.75$\pm$0.24 & 0.30$\pm$0.10 & 0.21$\pm$0.07\\
V986\,Oph        & r& 0840200501 & 2019-03-26T13:06:06  & 16.1 & B0IIIne  & 6.15 & 0.185 &1091$\pm$90 & 4.88$\pm$0.07 & 15.4$\pm$0.5  & 3.67$\pm$0.22 & 3.33$\pm$0.18\\
Sheliak          & r& 0840200901 & 2019-09-18T15:33:40  &  0.2 & B7Ve     & 3.52 & 0.033 & 296$\pm$15 & 3.82$\pm$0.04 & unobserved    & 12.2$\pm$3.0  & 11.2$\pm$2.1 \\
Sheliak          & r& 0840201501 & 2019-10-13T04:27:39  &  6.8 & B7Ve     & 3.52 & 0.033 & 296$\pm$15 & 3.82$\pm$0.04 & 49.7$\pm$1.3  & 12.2$\pm$0.6  & 12.3$\pm$0.5 \\
V558\,Lyr        & r& 0840200401 & 2019-10-03T04:18:22  &  7.7 & B3Ve     & 6.29 & 0.036 & 572$\pm$18 & 3.67$\pm$0.03 & 58.3$\pm$1.3  & 18.2$\pm$0.6  & 16.4$\pm$0.6 \\
59\,Cyg          & s& 0820310501 & 2018-05-17T11:10:57  & 12.7 & B1.5Vnne & 4.74 & 0.041 & 414$\pm$59 & 4.20$\pm$0.12 & 3.27$\pm$0.24 & 0.67$\pm$0.10 & 0.54$\pm$0.09\\
$\psi^{2}$\,Aqr  & r& 0820311101 & 2018-06-14T08:12:19  & 16.1 & B7.5     & 4.40 & 0.004 &  93$\pm$4  & 2.38$\pm$0.03 & 7.63$\pm$0.31 & 1.92$\pm$0.14 & 1.63$\pm$0.13\\
\hline      
\end{tabular}
\end{table*}

The \xmm\ data were processed with the Science Analysis Software (SAS) v18.0.0 using calibration files available in Oct. 2019 and following the recommendations of the \xmm\ team\footnote{SAS threads, see \\ http://xmm.esac.esa.int/sas/current/documentation/threads/ }. After the initial pipeline processing, the European Photon Imaging Camera (EPIC) observations were filtered to keep only the best-quality data ({\sc{pattern}} 0--12 for MOS and 0--4 for pn). To assess the crowding near the targets in order to choose the best extraction region, a source detection was performed on each EPIC dataset using the task {\it edetect\_chain}, which uses first sliding box algorithms and then performs a PSF fitting, on the 0.3--10.0\,keV energy band and for a log-likelihood of 10. We obtained a formal detection for all our targets but two, $\mu$\,Lup and $\kappa$\,Dra. The EPIC count rates of the others are provided in Table \ref{journal}. The detected X-ray sources lie at 2\arcsec\ or less for all stars (average separation is 1.2\arcsec) but three: d\,Lup is at 3.4\arcsec\ of its X-ray counterpart, I\,Hya at 3.7\arcsec, and HD\,43285 at 4.2\arcsec. Contrary to catalogs like the 3XMM, no further astrometric correction has been applied hence small shifts remain possible. Unfortunately, the lack of X-ray sources in the I\,Hya and HD\,43285 fields prevents us to check source alignment for other targets and therefore the accuracy of the astrometry calculation of the X-ray counterparts which are associated with our targets. Near d\,Lup, however, star 2MASS J15355876--4456355 is detected and it appears at 1.4\arcsec\ only of the X-ray source. We therefore consider these three detections as tentative, awaiting independent confirmation. For three targets with low count rates (QY\,Car, d\,Lup, $\alpha$\,Ara), it was not possible to extract a spectrum hence we estimated the source hardness by performing a second detection run, this time using the 0.5--2.0\,keV and 2.0--10.0\,keV energy bands. Finally, it should be noted that the observation of HD\,120678 was affected by straylight from a nearby bright source (probably the RSCVn HD\,119285).

Light curves for events beyond 10\,keV were built for the full fields. Using them, background flares were detected in about half of the exposures and the time intervals corresponding to flaring events were discarded before further processing. For the brighter detections (i.e. EPIC-pn count rate larger than 0.01\,cts\,s$^{-1}$), we then extracted EPIC light curves (in the 0.3--10.\,keV energy band) and spectra of the source using circular regions centered on the Simbad positions of the targets with radii between 12.5 and 30\arcsec, depending on the crowding and position of CCD gaps. Background was derived in nearby circular regions of 30\arcsec\ radius devoid of sources, except for HD\,120678 where nearby boxes were rather used for MOS as they better allow to avoid straylight contamination. Light Curves were corrected using the task {\it epiclccorr} to provide full-PSF, equivalent on-axis count rates. Time bins between 100s and 2000s were used, depending on source brightness, and bins with fractional exposure times smaller than 50\% were discarded. Spectra and their dedicated calibration matrices (ancillary response file and redistribution matrix file response matrices, which are used to calibrate the flux and energy axes, respectively) were derived using the task {\it{especget}}. EPIC spectra were grouped with {\it{specgroup}} to obtain an oversampling factor of five and to ensure that a minimum signal-to-noise ratio of 3 (i.e., a minimum of ten counts) was reached in each spectral bin of the background-corrected spectra. 

\section{Results}

\begin{table*}
\centering
\caption{Results of the spectral fitting (see text for details). }
\label{fits}
\tiny
\setlength{\tabcolsep}{1pt}
\begin{tabular}{lccccccccccccc}
\hline\hline
Name                    & $N_{\rm H}^{ISM}$ &$N_{\rm H}$ & $kT_1$ & $norm_1$ &$ kT_2$ & $norm_2$ & $kT_3$ & $norm_3$ & $\chi^2_{\nu}$(dof) & $F_{\rm X}^{obs}$(tot$^a$) & $L_{\rm X}^{ISM\,cor}$(tot) & $\log(L_{\rm X}^{ISM\,cor}(tot)$ & $HR^b$ \\
                        & \multicolumn{2}{c}{($10^{22}$\,cm$^{-2}$)} & (keV) & (cm$^{-5}$) & (keV) & (cm$^{-5}$) & (keV) & (cm$^{-5}$) & & (erg\,cm$^{-2}$\,s$^{-1}$) & (erg\,s$^{-1}$) & $/L_{\rm BOL})$ & \\
\hline
HD\,18552       & 0.024 &0.027$\pm$0.013 & 0.23$\pm$0.02  & (7.57$\pm$1.26)e-5 & 0.93$\pm$0.05 & (8.04$\pm$1.17)e-5 & 2.08$\pm$0.51 & (8.53$\pm$1.49)e-5 & 1.20(136) & (3.05$\pm$0.11)e-13 & (1.98$\pm$0.13)e30 & --5.767$\pm$0.016 & 0.20$\pm$0.04     \\
$\eta$\,Ori     & 0.004 &0.032$\pm$0.008 &0.187$\pm$0.003 & (4.55$\pm$0.36)e-4 &               &                    &               &                    & 1.66(73)  & (2.78$\pm$0.10)e-13 & (3.72$\pm$2.39)e30 & --7.653$\pm$0.016 & (1.74$\pm$0.23)e-4\\
HR\,1847        & 0.094 & 0.00$\pm$0.04  & 0.89$\pm$0.07  & (1.22$\pm$0.16)e-5 &               &                    &               &                    & 2.44(18)  & (2.14$\pm$0.32)e-14 & (5.82$\pm$1.01)e29 & --7.119$\pm$0.065 & 0.07$\pm$0.06     \\
HD\,43285       & 0.010 & 0.08$\pm$0.05  &0.252$\pm$0.018 & (6.86$\pm$2.49)e-5 & 1.28$\pm$0.08 & (3.67$\pm$0.45)e-5 &               &                    & 1.10(48)  & (9.84$\pm$0.79)e-14 & (6.01$\pm$0.53)e29 & --6.584$\pm$0.035 & 0.10$\pm$0.02     \\
HD\,44458       & 0.065 &0.096$\pm$0.010 & 0.98$\pm$0.05  & (6.81$\pm$0.88)e-5 & 7.27$\pm$0.72 & (1.09$\pm$0.02)e-3 &               &                    & 1.00(394) & (1.87$\pm$0.03)e-12 & (8.90$\pm$1.22)e31 & --6.024$\pm$0.007 & 2.08$\pm$0.06     \\
HD\,45995       & 0.063 &0.027$\pm$0.018 & 1.28$\pm$0.34  & (5.45$\pm$3.19)e-5 & 7.34$\pm$1.73 & (5.78$\pm$0.47)e-4 &               &                    & 0.98(156) & (1.05$\pm$0.07)e-12 & (5.72$\pm$0.85)e31 & --5.856$\pm$0.029 & 1.86$\pm$0.16     \\
I\,Hya          & 0.009 &0.000$\pm$0.002 & 2.35$\pm$0.08  & (4.17$\pm$0.07)e-4 &               &                    &               &                    & 1.91(223) & (5.24$\pm$0.13)e-13 & (1.48$\pm$0.14)e30 & --6.322$\pm$0.011 & 0.75$\pm$0.04     \\
HD\,120678      & 0.26  & 0. (fixed)     & 0.77$\pm$0.06  & (1.04$\pm$0.14)e-5 & 8.83$\pm$5.77 & (5.34$\pm$0.33)e-5 &               &                    & 1.16(71)  & (9.95$\pm$1.49)e-14 & (7.89$\pm$1.88)e31 & --6.786$\pm$0.065 & 1.34$\pm$0.32     \\
V986\,Oph       & 0.11  & 0.08$\pm$0.04  & 0.16$\pm$0.02  & (3.07$\pm$1.45)e-4 & 0.73$\pm$0.03 & (7.24$\pm$0.88)e-5 &               &                    & 1.49(79)  & (1.65$\pm$0.14)e-13 & (3.40$\pm$0.63)e31 & --6.931$\pm$0.037 & 0.025$\pm$0.003   \\
Sheliak$^c$     & 0.020 & 0.37$\pm$0.04  &0.043$\pm$0.005 &  5.09$\pm$6.74     &0.571$\pm$0.018& (6.65$\pm$0.66)e-4 & 24.2$\pm$16.6 & (1.07$\pm$0.29)e-4 & 1.19(147) & (8.03$\pm$1.86)e-13 & (8.78$\pm$2.22)e30 & --6.460$\pm$0.100 & 0.25$\pm$0.21     \\
V558\,Lyr       & 0.022 &0.083$\pm$0.011 & 1.30$\pm$0.07  & (1.37$\pm$0.27)e-4 & 12.8$\pm$2.95 & (1.00$\pm$0.04)e-3 &               &                    & 1.01(237) & (1.93$\pm$0.07)e-12 & (7.64$\pm$0.55)e31 & --5.370$\pm$0.016 & 2.31$\pm$0.13     \\
59\,Cyg         & 0.025 &0.018$\pm$0.022 &0.249$\pm$0.014 & (4.24$\pm$0.71)e-5 &               &                    &               &                    & 1.67(25)  & (3.67$\pm$0.40)e-14 & (8.27$\pm$2.52)e29 & --7.866$\pm$0.047 & (9.85$\pm$2.46)e-4\\
$\psi^{2}$\,Aqr & 0.002 &0.042$\pm$0.014 &0.231$\pm$0.017 & (5.24$\pm$0.73)e-5 & 0.99$\pm$0.04 & (3.44$\pm$0.25)e-5 &               &                    & 1.07(73)  & (1.03$\pm$0.04)e-13 & (1.08$\pm$0.10)e29 & --6.931$\pm$0.017 & 0.059$\pm$0.007   \\
\hline
\multicolumn{14}{l}{$^a$ The total band (tot) corresponds to the 0.5--10.0\,keV energy band.}\\
\multicolumn{14}{l}{$^b$ $HR$ is defined, as in \citet{naz18}, as the ratio between the ISM-corrected fluxes in the hard (2.0--10.0\,keV) and soft (0.5--2.0\,keV) energy bands. }\\
\multicolumn{14}{l}{$^c$ This fit refers to ObsID 0840201501, the other exposure being too short for more than a simple detection.}\\
\end{tabular}
\end{table*}

Our sample consisted of 18 Be stars. Two of these, $\mu$\,Lup and $\kappa$\,Dra, are not detected in the new observations. The previous claimed {\it ROSAT} detection of $\mu$\,Lup can however be easily explained by confusion as an X-ray source actually appears at 22\arcsec\ southeast of the target. It corresponds to the A-star HD\,135748. The previous {\it Exosat} detection of $\kappa$\,Dra, on the other hand, is most probably due to optical/UV loading from this bright hot star.

\begin{figure}
  \begin{center}
\includegraphics[width=8cm]{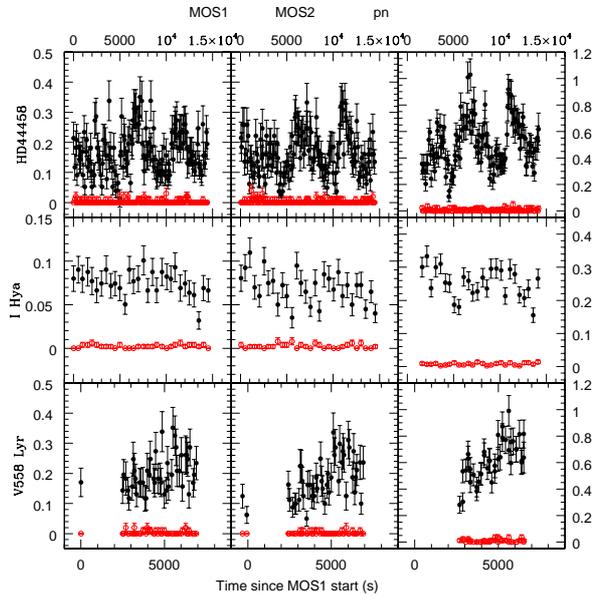}
  \end{center}
  \caption{Background-corrected EPIC light curves in 0.3--10.\,keV for HD\,44458 (top), I\,Hya (middle), and V558\,Lyr (bottom), with MOS1 on the left, MOS2 in the middle, and pn on the right. Open red symbols correspond to the background light curves; x-axes are the same for HD\,44458 and I\,Hya and, for each star, y-axes are the same for both MOS.}
\label{lc}
\end{figure}

Three further targets, QY\,Car, d\,Lup, and $\alpha$\,Ara, are detected but remain faint in \xmm\ data. A detection run in two energy bands (soft, 0.5--2.\,keV, and hard, 2.--10.\,keV) reveals that the soft count rate is at least three times larger than the hard one, clearly indicating that these stars are emitting mostly at low energies. To estimate their X-ray luminosities, we converted their count rates within WebPIMMs\footnote{https://heasarc.gsfc.nasa.gov/cgi-bin/Tools/w3pimms/w3pimms.pl} using temperatures of 0.3 or 1\,keV (since we know their softness but without a precise temperature constraint) and the interstellar absorption derived from the color excess (Table \ref{journal}) using the relation of \citet{gud12}. The resulting luminosities are 0.7--40., 1.3--2.7, 0.6--2.3$\times 10^{28}$\,erg\,cm$^{-2}$\,s$^{-1}$ for QY\,Car, d\,Lup, and $\alpha$\,Ara, respectively. This corresponds to $\log(L_{\rm X}/L_{\rm BOL})$ of --9.6..--7.8, --8.6..--8.3, and --9.2..--8.6 for QY\,Car, d\,Lup, and $\alpha$\,Ara, respectively. Clearly, those three stars do not display the typical \gc\ characteristics.

The light curves and spectra of the 13 remaining targets can be analyzed in some detail. Using $\chi^2$ tests, significant ($SL<1$\%) variability is detected for the background-corrected EPIC light curves of HD\,44458, I\,Hya, and V558\,Lyr (Fig. \ref{lc}). Moreover, since most of the targets had been previously detected in the X-ray range, we also investigated the longer-term variability. By folding the best-fit \xmm\ spectral model (see text below and Table \ref{fits}) through the {\it ROSAT} response matrices, we predicted an equivalent {\it ROSAT} count rate (Table \ref{ros}), which we compared to the one reported by \citet{ber96}. The differences between predicted and observed count rates are always below 3$\sigma$; in fact, they are even below 2$\sigma$ for all but one source (HD\,44458, 2.4$\sigma$) which may hint at long-term changes for that star. For d\,Lup, \xmm\ spectra could not be extracted hence the \xmm\ count rate was transformed into its {\it ROSAT} equivalent within WebPIMMs, with the same hypotheses as above. The predicted count rate is significantly (3.3$\sigma$) below the observed {\it ROSAT} value, implying a decrease of the X-ray emission. For {\it Exosat}-ME, similarly converting the \xmm\ count rates of QY\,Car leads to very small values ($<2\times 10^{-4}$\,ct\,s$^{-1}$), strongly suggesting that the previous detection could have been spurious, i.e. due to optical/UV loading (as for $\kappa$\,Dra). In parallel, $\alpha$\,Ara is reported in the XMM slew survey catalog v2.0 as XMMSL2 J173150.8--495233 with an EPIC-pn count rate of 2.1$\pm$0.9\,ct\,s$^{-1}$ in the 0.2--12.\,keV energy band. This is fully compatible with our value in the 0.3--10.\,keV band (Table \ref{journal}) but the slew survey error is very large and only permits to detect drastic changes of brightness. $\alpha$\,Ara was however observed twice by \xmm\ in pointed observations and these two exposures display differences larger than 3$\sigma$ for the EPIC-pn count rate (for the less sensitive MOS cameras, the rate differences amount to 1--2$\sigma$). This implies that this star undergoes variations of its X-ray flux on timescales of months. Finally, the two \xmm\ observations of Sheliak provide similar count rates, suggesting a rather constant X-ray emission from that star.

\begin{table}
\centering
\caption{{\it ROSAT} observed count rates \citep{ber96} and expected values assuming the XMM fluxes or count rates (see text for details).}
\label{ros}
\setlength{\tabcolsep}{3.3pt}
\begin{tabular}{ccc}
  \hline\hline
  Name & Obs & Pred \\
       & \multicolumn{2}{c}{(ct\,s$^{-1}$)}\\
  \hline
HD\,18552        & 0.017$\pm$0.008 & 0.031\\
$\eta$\,Ori      & 0.063$\pm$0.016 & 0.054\\
HR\,1847         & 0.009$\pm$0.006 & 0.002\\
HD\,43285        & 0.017$\pm$0.007 & 0.010\\
HD\,44458        & 0.030$\pm$0.009 & 0.052\\
HD\,45995        & 0.024$\pm$0.010 & 0.032\\
I\,Hya           & 0.035$\pm$0.011 & 0.043\\
d\,Lup           & 0.054$\pm$0.016 & 0.001\\
V986\,Oph        & 0.020$\pm$0.011 & 0.016\\
Sheliak          & 0.073$\pm$0.011 & 0.068\\
V558\,Lyr        & 0.042$\pm$0.010 & 0.053\\
$\psi^{2}$\,Aqr  & 0.021$\pm$0.010 & 0.014\\
\hline      
\end{tabular}
\end{table}

\begin{figure*}
  \begin{center}
\includegraphics[width=5.8cm]{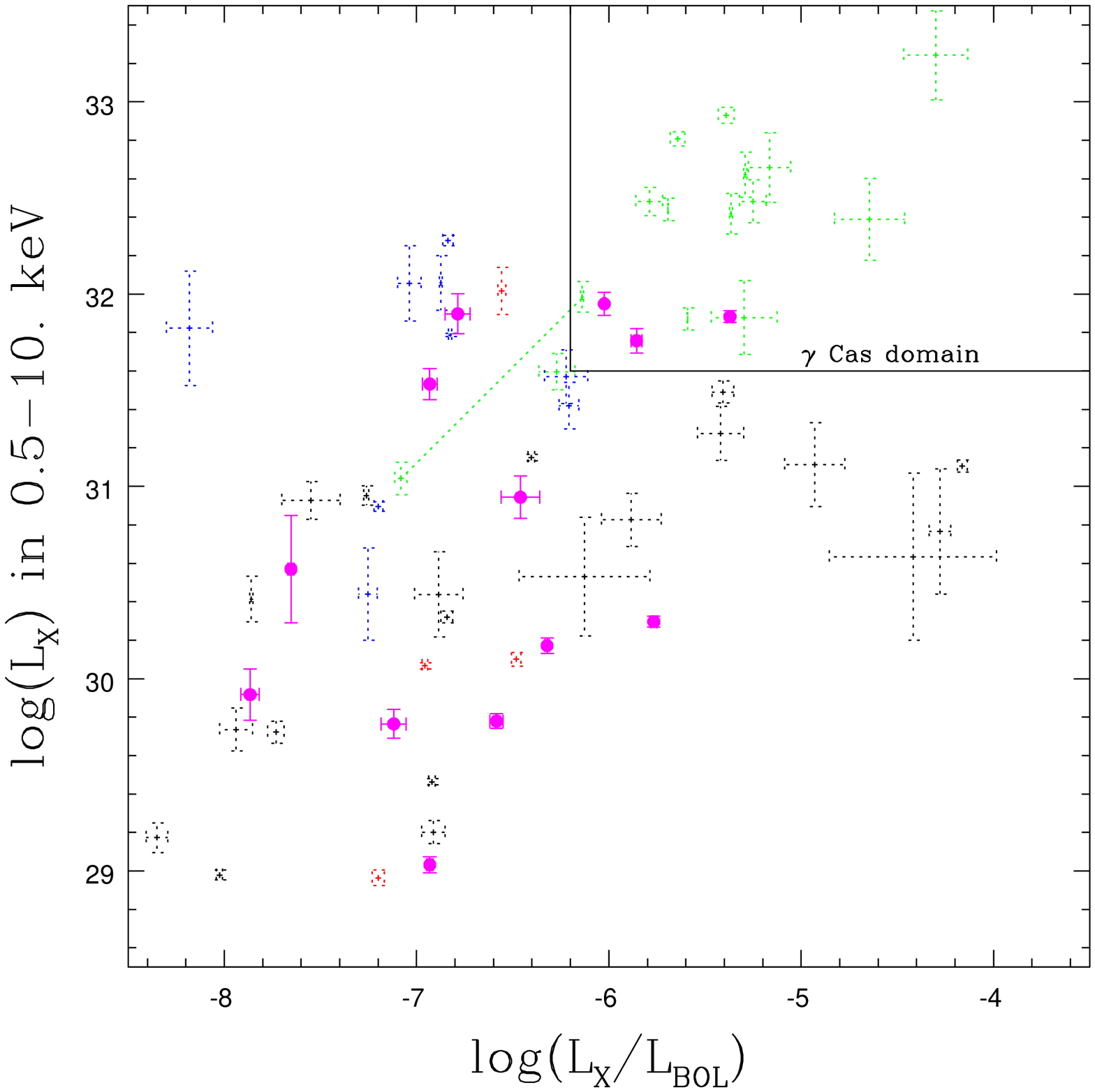}
\includegraphics[width=5.8cm]{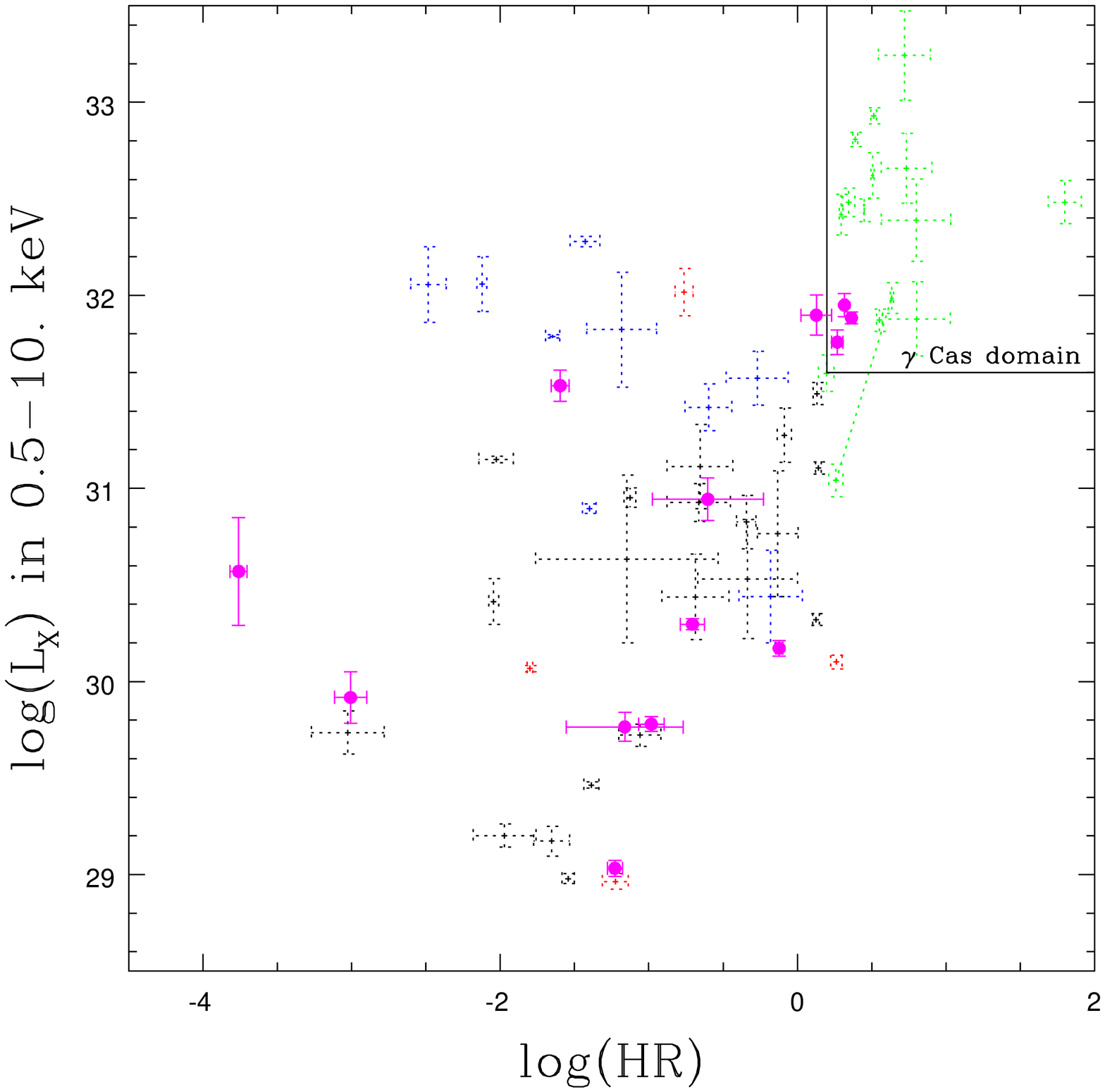}
\includegraphics[width=5.8cm]{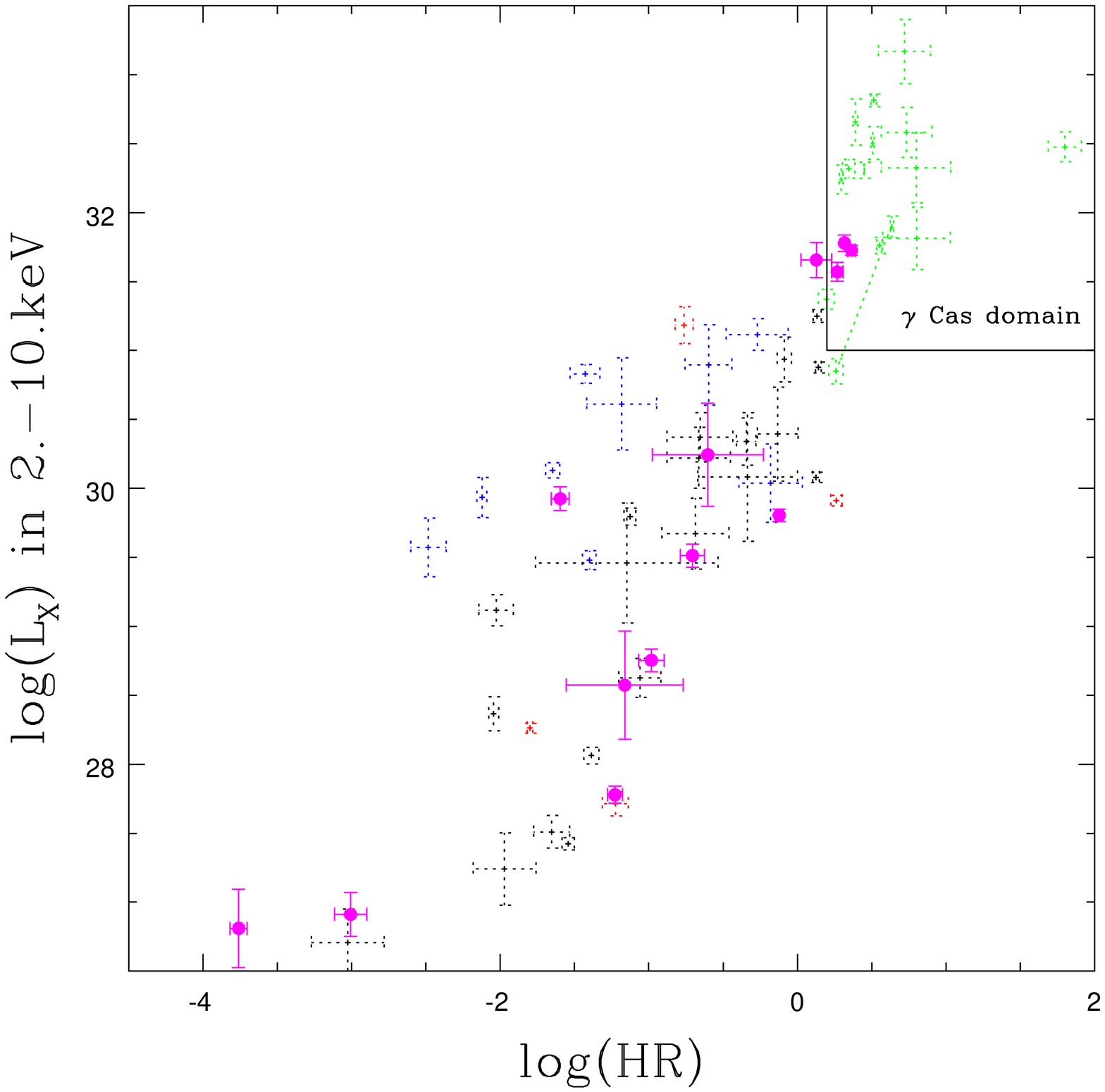}
  \end{center}
  \caption{Comparison of X-ray luminosities (either in total, 0.5--10.\,keV, band or in hard, 2.--10.\,keV, band) with hardness ratios $HR=Hard/Soft$ or X-ray to bolometric luminosity ratio $L_{\rm X}/L_{\rm BOL}$. Dotted lines are used for the sample of \citet{naz18}: green for \gc\ stars (the two recorded states of HD\,45314 are linked by a straight line), blue for non-magnetic O-stars, red for magnetic objects, and black for other stars. Magenta symbols correspond to targets of this paper. Proposed limits on the \gc\ domain are drawn with solid black lines (in this context, note that no Be-XRBs are present within our sample hence they are absent from these plots).}
\label{compa}
\end{figure*}

All available EPIC spectra were fitted simultaneously within Xspec v12.9.1p, as in \citet{naz18}. As is usual for massive stars, we used absorbed optically thin thermal plasma models (i.e., $tbabs\times phabs\times \sum apec$) with solar abundances of \citet{asp09}. The first absorption component was fixed to the interstellar column, derived from the known color excess (Table \ref{journal}) using the formula of \citet[$N_{\rm H}^{ISM}=6.12\times 10^{21}\times E(B-V)$\,cm$^{-2}$]{gud12}, whereas the second absorbing component accounts for possible local absorption and was allowed to vary. In one case, however, the additional absorption converged to $\sim$0 and yielded erratic results with unrealistic errors, hence we fixed it to zero. For the emission components, we used up to three temperatures, depending on the goodness of fit. Final fitting results are provided in Table \ref{fits}. Errors correspond to 1$\sigma$ uncertainties; whenever they were asymetric, the largest value is reported in Table \ref{fits}. For the X-ray luminosities (in total band 0.5--10.\,keV), errors combine the distance errors (see Table \ref{journal}) with errors on X-ray fluxes (derived from the ``flux err'' command in Xspec), but do not integrate the impact of model choices; errors on $\log(L_{\rm X}/L_{\rm BOL})$, however, do not depend on distance and reflect only X-ray flux uncertainties. Hardness ratios $HR$ were calculated as the ratios between the fluxes, corrected for interstellar absorption, in the hard (2.0--10.0 keV) and soft (0.5--2.0 keV) energy bands.

Figure \ref{compa} compares the X-ray luminosities, bolometric luminosities, and hardness of the X-ray emission of our 13 brightest targets to the sample of \citet{naz18}. It is immediately obvious that the new X-ray sources lie amongst the previous sample of Be stars. To assess whether new \gc\ analogs are amongst them, we recall the criteria for such a classification: presence of the K$\alpha$ fluorescence line from relatively low-ionization iron near the high-ionization iron lines at 6.7-7.0\,keV (only visible in well exposed spectra), presence of variability (on short and/or long-term), large but not extreme X-ray brightness ($\log(L_{\rm X}^{ISM\,cor}(0.5-10\,{\rm keV})\sim 31.6-33.2$ or $\log(L_{\rm X}/L_{\rm BOL})$ between --6.2 and --4.), unusual hardness ($kT>5.$\,keV, $HR>1.6$, $L_{\rm X}^{ISM\,cor}(2.-10\,{\rm keV})>10^{31}$\,erg\,cm$^{-2}$\,s$^{-1}$).

\begin{figure}
  \begin{center}
\includegraphics[width=8cm]{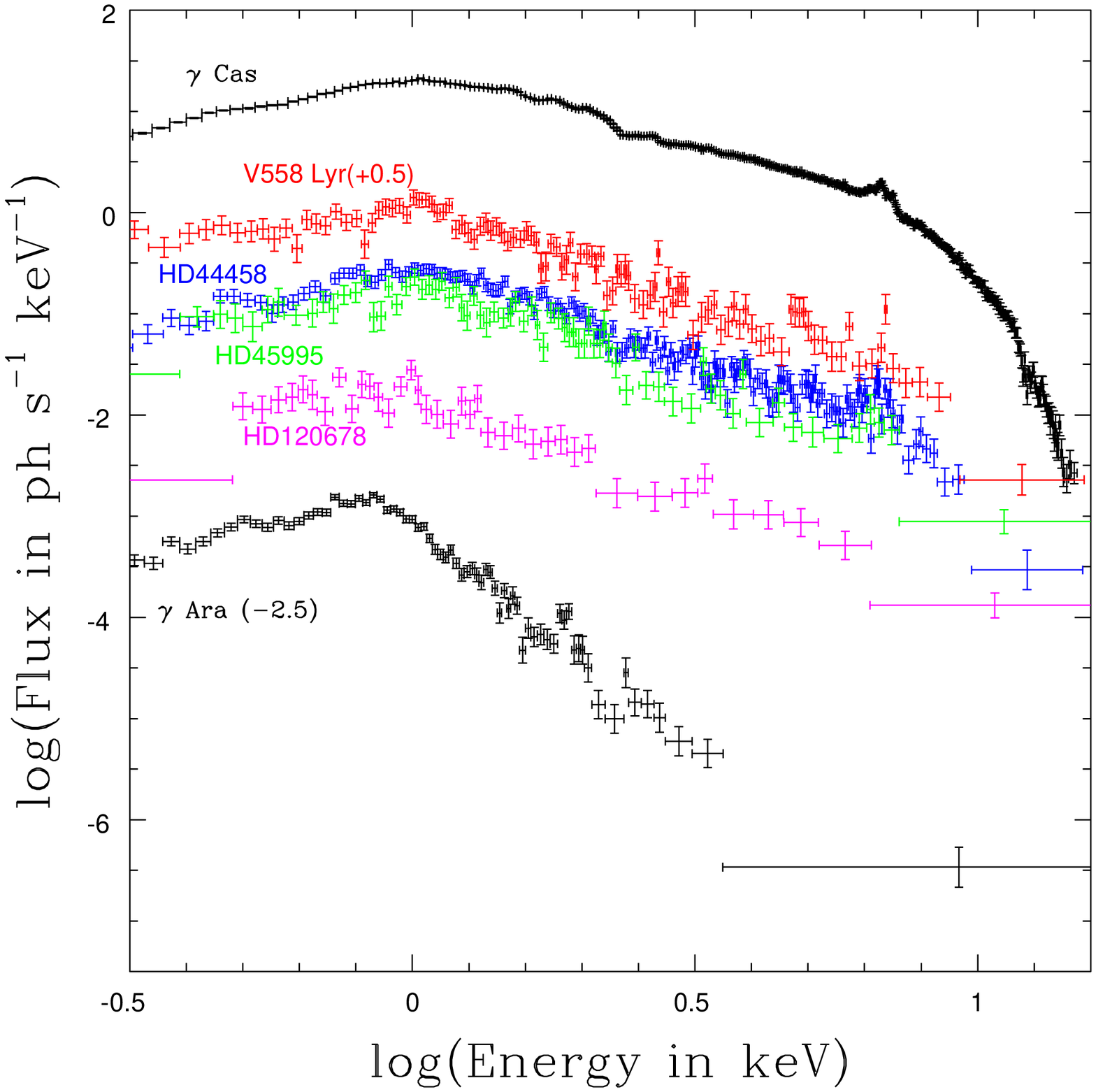}
\includegraphics[width=8cm]{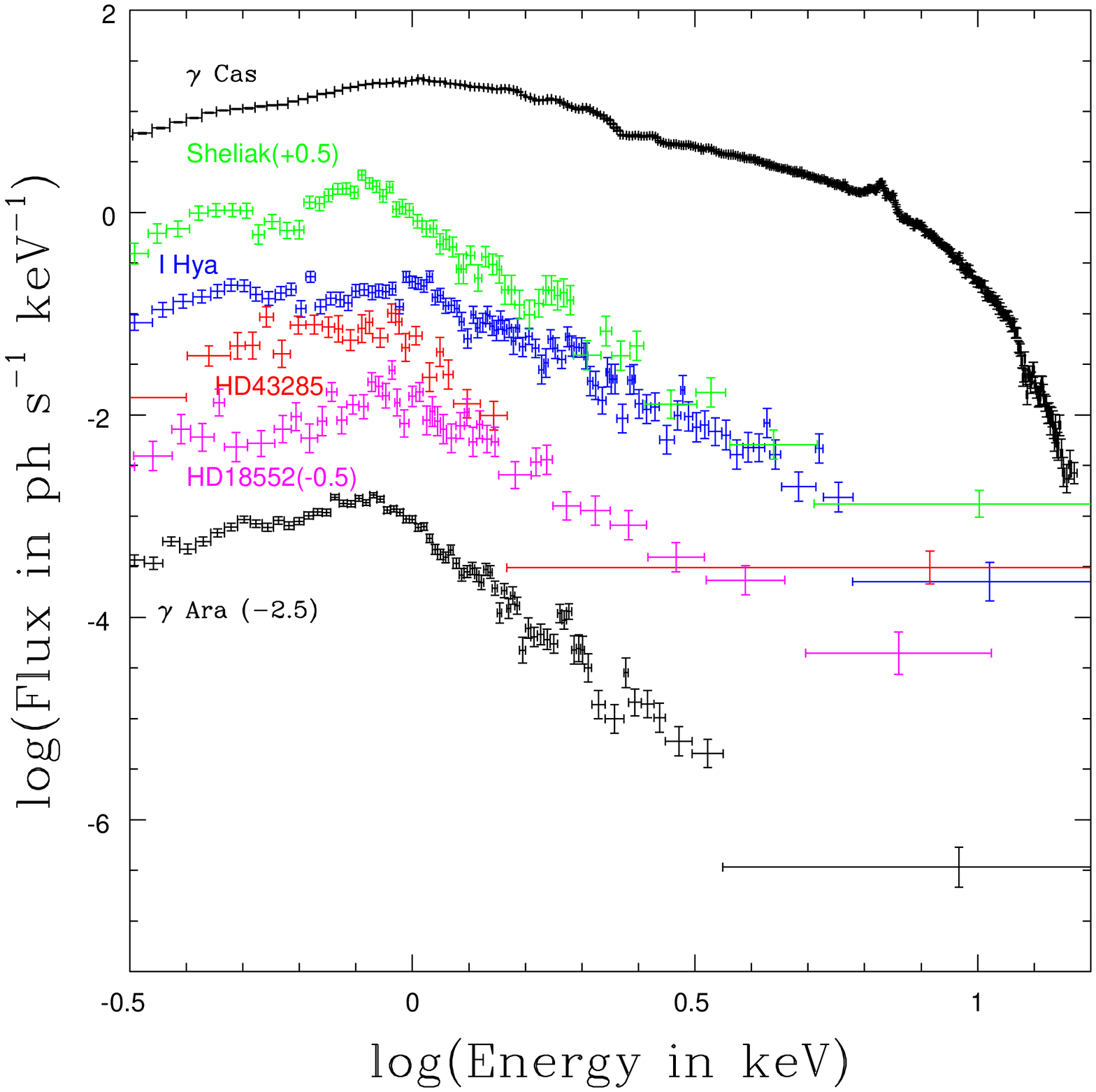}
  \end{center}
  \caption{X-ray spectra of targets with a hard spectrum (top: new \gc\ stars and candidate) and a slightly hard spectrum (bottom), compared to those of \gc\ and of the ``normal'' X-ray emission from the Be-star $\gamma$\,Ara (B1IIe). Arbitrary vertical shifts were sometimes applied to facilitate comparison, they are quoted between brackets after the star's name. Those spectra were taken with EPIC-pn, except for HD\,18552 (MOS1).}
\label{xspec}
\end{figure}

Amongst the 13 targets, four display an intense X-ray emission at high energies ($L_{\rm X}^{ISM\,cor}(2.-10\,{\rm keV})=3-6\times 10^{31}$\,erg\,cm$^{-2}$\,s$^{-1}$): HD\,44458, HD\,45995, V558\,Lyr, and HD\,120678. The first three stars also display large hardness ratios ($HR=1.9-2.3$, $kT=7-13$\,keV) and rather intense overall X-ray emission ($L_{\rm X}^{ISM\,cor}(0.5-10\,{\rm keV})=6-9\times 10^{31}$\,erg\,cm$^{-2}$\,s$^{-1}$, $\log(L_{\rm X}/L_{\rm BOL})=-6.0$ to $-5.4$). Those values are well beyond what could be expected from a PMS companion. Furthermore, HD\,44458 and V558\,Lyr were found to be significantly variable and the spectra of HD\,44458 and HD\,45995 clearly show the presence of the iron complex (Fig. \ref{xspec}). As a complement, the BeSS database was searched for optical spectra taken close to the \xmm\ observation dates of those three stars. High-resolution TIGRE spectra of HD\,44458 and HD\,45995 were also taken in the framework of a stellar survey of B-stars to be observed with {\it eROSITA} (PI: J. Robrade). While not strictly simultaneous, as they were obtained some months before or after the \xmm\ observations, these spectra provide an idea of the disk emission strength, which is usually quite high in \gc\ stars. These spectra were corrected within IRAF from telluric absorptions using the template of \citet{hin00} and then normalized using low-order polynomials. The equivalent width ($EW$) of the H$\alpha$ line was estimated from --540\,km\,s$^{-1}$ to +540\,km\,s$^{-1}$ and its values are reported in Table \ref{ha}. In all three cases, a dense disk seems to be present. As a last information, we may add that the three stars display quite early spectral types, again a typical feature of \gc\ stars. Therefore, HD\,44458, HD\,45995, and V558\,Lyr clearly display \gc\ characteristics and can be added to the list of such objects.

\begin{table}
\centering
\caption{$EW$ of the H$\alpha$ line measured on optical spectra (see text for details). The labels ``b,t'' indicate the source of the data (BeSS or TIGRE, respectively).}
\label{ha}
\setlength{\tabcolsep}{3.3pt}
\begin{tabular}{cccc}
  \hline\hline
  Name & Date & $EW$ (\AA) & \\
  \hline
HD\,44458        & 2017-10-18& --37.6& b\\
        & 2018-11-25& --38.4& t\\
        & 2018-12-11& --37.4& t\\
        & 2019-02-05& --35.6& b\\
        & 2019-02-09& --37.0& t\\
        & 2019-02-15& --35.8& b\\
        & 2019-03-05& --30.2& t\\
\vspace*{-0.2cm}\\
HD\,45995        & 2018-03-22& --16.8& b\\
        & 2018-11-24& --21.3& t\\
        & 2018-12-10& --21.4& t\\
        & 2019-01-19& --12.9& b\\
        & 2019-02-05& --22.0& t\\
        & 2019-03-03& --22.6& t\\
\vspace*{-0.2cm}\\
V558\,Lyr        & 2019-08-01& --23.9& b\\
\hline      
\end{tabular}
\end{table}

HD\,120678, which had undergone a shell-event in mid-2008 \citep{gam12}, appears slightly less bright (its $\log(L_{\rm X}/L_{\rm BOL})$ only amounts to --6.8) and less hard considering its $HR$ of 1.3, though its spectrum still requires a $\sim$9\,keV component to be fitted and its hard X-ray emission is well above $10^{31}$\,erg\,cm$^{-2}$\,s$^{-1}$. However, its X-ray observation was strongly affected by straylight, even if that contamination remains minimal at the source position. An independent confirmation of its X-ray properties would thus be welcome and, until then, we will consider it as a \gc\ candidate.

\begin{table}
\centering
\caption{List of all \gc\ analogs known to date (\citealt{smi16} and references therein, \citealt{naz18}, and this paper).}
\label{gc}
\setlength{\tabcolsep}{3.3pt}
\begin{tabular}{ccc}
\hline\hline
\multicolumn{3}{l}{\gc\ stars}\\
\gc\ & TYC\,3681-695-1 & V782\,Cas\\
HD\,44458 & HD\,45314 & HD\,45995 \\
HD\,90563 & HD\,110432 & HD\,119682 \\
V767\,Cen & CQ\,Cir & HD\,157832\\
HD\,161103 & V771\,Sgr & HD\,316568\\
2XMMJ\,180816.6-191939 & GSC2\,S300302371 & SS397\\
Cl*\,NGC\,6649\,WL\,9 & 3XMMJ\,190144.5+045914 & V558\,Lyr\\
SAO\,49725 & V2156\,Cyg & $\pi$\,Aqr\\
V810\,Cas\\
\vspace*{-0.2cm}\\
\multicolumn{3}{l}{\gc\ candidates}\\
HD\,42054 & HD\,120678\\
\hline      
\end{tabular}
\end{table}

Finally, Sheliak ($\beta$\,Lyr) does seem slightly hard ($HR=0.25$) and quite bright compared to its bolometric luminosity ($\log(L_{\rm X}/L_{\rm BOL})=-6.5$). Its high-energy tail requires three thermal components for achieving a good fit, and the third temperature is $kT\sim24$\,keV. Those are very unusual features for a B7 star and their possible relationship with the companion's presence remains to be explored, hence we consider it as an interesting target deserving further study. A few other stars appear to display a slightly hard spectrum, though not with the extreme characteristics of \gc\ stars: HD\,18552 has $HR=0.2$ and $\log(L_{\rm X}/L_{\rm BOL})=-5.8$; HD\,43285 shows $HR=0.1$ and $\log(L_{\rm X}/L_{\rm BOL})=-6.6$; I\,Hya is variable and has $HR=0.75$ with $\log(L_{\rm X}/L_{\rm BOL})=-6.3$. We however recall that HD\,43285 and I\,Hya appeared somewhat distant from their \xmm\ counterpart, casting some doubt on their association (see Section 2). Figure \ref{xspec} compares the spectra of these sources to those of \gc\ and of a ``normal'' massive star: a hard tail, though steeper than for \gc\ stars, is clearly present. All other stars are much less bright, with $L_{\rm X}^{ISM\,cor}(0.5-10\,{\rm keV})<4\times 10^{31}$\,erg\,cm$^{-2}$\,s$^{-1}$, and less hard ($HR<0.1$).

\section{Summary and conclusions}
In this paper, we continue our search for \gc\ analogs thanks to a small dedicated X-ray survey. \xmm\ was pointed at 18 Be stars with previous reports of X-ray detections ({\it ROSAT, Exosat, XMM}-slew survey) or which underwent a ``shell'' event as did the \gc\ star HD\,45314. Two of these stars ($\mu$\,Lup and $\kappa$\,Dra) remain undetected and three others (QY\,Car, d\,Lup, $\alpha$\,Ara) display only a faint and soft X-ray emission leading to a simple detection. The remaining 13 targets could be studied spectroscopically. Amongst the detections, three X-ray sources appear at distances of 3--4\arcsec\ from the locations of their Be counterparts (d\,Lup, I\,Hya, and HD\,43285), hence their associations are only considered as tentative.

Both the short-term and long-term variability of the sources were examined. The EPIC light curves of the 13 X-ray brightest targets were extracted and tested against constancy using $\chi^2$ tests: only HD\,44458, I\,Hya, and V558\,Lyr appear significantly variable over an exposure. Turning to the long-term behaviour, we find that the previous {\it ROSAT} detections and \xmm\ observations generally agree well, except for a 2.4$\sigma$ change in HD\,44458 and a 3.3$\sigma$ difference in d\,Lup. The analysis of pairs of \xmm\ observations further indicates a significant (3$\sigma$) change on a five-months timescale in $\alpha$\,Ara. Finally, because the X-ray brightnesses found by \xmm\ are very low, the previous {\it Exosat} detections of $\kappa$\,Dra and QY\,Car were probably due to UV contamination.

The X-ray properties of the 13 X-ray brightest targets appear in line with those found in a general survey of Be stars. With their hard and moderately bright X-ray emission, three stars clearly belong to the \gc\ category: HD\,44458, HD\,45995, and V558\,Lyr. An additional one, HD\,120678, can be considered as a \gc\ candidate as it is only slightly softer. It may be noted that the new detections occurred at the low-luminosity end amongst the \gc\ cases, while the hardness of their spectra clearly sets them apart from the typical massive stars. Furthermore, the studied Be stars seem to show a continuum of behaviours in luminosity/hardness plots, suggesting that, perhaps, low-level \gc\ activity is possible.

This study brings the total of known \gc\ objects to 25, with two \gc\ candidates (Table \ref{gc}). In just two years, the number of such objects has doubled. This shows that the \gc\ phenomenon is not a rare peculiarity but that such stars constitute a true new class of astronomical objects, whose exact nature remains to be determined, however. Future surveys, such as that performed by {\it e-ROSITA} will certainly find additional cases and constrain their actual incidence rate in distance-limited samples.

\section*{Acknowledgements}
Y.N., G.R., and S.K. acknowledge support from the Fonds National de la Recherche Scientifique (Belgium), the European Space Agency (ESA) and the Belgian Federal Science Policy Office (BELSPO) in the framework of the PRODEX Programme (contract XMaS). J.M.T. acknowledges support from the research grant  ESP2017-85691-P. J.R. acknowledges support from DLR under grant 50OR1605. ADS and CDS were used for preparing this document. This work has also made use of the BeSS database, operated at LESIA (Observatoire de Meudon, France) and available on http://basebe.obspm.fr

\bsp	
\label{lastpage}

\begin{thebibliography}{99}
\bibitem[\protect\citeauthoryear{Asplund et al.}{2009}]{asp09} Asplund, M., Grevesse, N., Sauval, A.J., \& Scott, P.\ 2009, \araa, 47, 481 
\bibitem[\protect\citeauthoryear{Babel \& Montmerle}{1997}]{bab97} Babel, J., \& Montmerle, T.\ 1997, \aap, 323, 121
\bibitem[\protect\citeauthoryear{Bailer-Jones et al.}{2018}]{bai18} Bailer-Jones, C.~A.~L., Rybizki, J., Fouesneau, M., Mantelet, G., \& Andrae, R.\ 2018, \aj, 156, 58
\bibitem[\protect\citeauthoryear{Berghoefer et al.}{1996}]{ber96} Berghoefer, T.~W., Schmitt, J.~H.~M.~M., \& Cassinelli, J.~P.\ 1996, \aaps, 118, 481
\bibitem[\protect\citeauthoryear{Berghoefer et al.}{1997}]{ber97} Berghoefer, T.~W., Schmitt, J.~H.~M.~M., Danner, R., \& Cassinelli, J.~P.\ 1997, \aap, 322, 167 
\bibitem[\protect\citeauthoryear{Capitanio et al.}{2017}]{cap17} Capitanio, L., Lallement, R., Vergely, J.~L., Elyajouri, M., \& Monreal-Ibero, A.\ 2017, \aap, 606, A65 
\bibitem[\protect\citeauthoryear{Cantiello \& Braithwaite}{2011}]{can11} Cantiello, M., \& Braithwaite, J.\ 2011, \aap, 534, A140
\bibitem[\protect\citeauthoryear{Drimmel et al.}{2019}]{dri19} Drimmel, R., Bucciarelli, B., \& Inno, L.\ 2019, Research Notes of the American Astronomical Society, 3, 79
\bibitem[\protect\citeauthoryear{Gamen et al.}{2012}]{gam12} Gamen, R., Arias, J.~I., Barb{\'a}, R.~H., et al.\ 2012, \aap, 546, A92
\bibitem[\protect\citeauthoryear{Grunhut et al.}{2012}]{gru12} Grunhut, J.~H., Wade, G.~A., \& MiMeS Collaboration 2012, American Institute of Physics Conference Series, 1429, 67 
\bibitem[\protect\citeauthoryear{Gudennavar et al.}{2012}]{gud12} Gudennavar, S.~B., Bubbly, S.~G., Preethi, K., \& Murthy, J.\ 2012, \apjs, 199, 8 
\bibitem[\protect\citeauthoryear{Hamaguchi et al.}{2016}]{ham16} Hamaguchi, K., Oskinova, L., Russell, C.~M.~P., et al.\ 2016, \apj, 832, 140
\bibitem[\protect\citeauthoryear{Hinkle et al.}{2000}]{hin00} Hinkle, K., Wallace, L., Valenti, J., \& Harmer, D.\ 2000, Visible and Near Infrared Atlas of the Arcturus Spectrum 3727-9300 A ed.~Kenneth Hinkle, Lloyd Wallace, Jeff Valenti, and Dianne Harmer.~(San Francisco: ASP) ISBN: 1-58381-037-4, 2000.,
\bibitem[\protect\citeauthoryear{Murakami et al.}{1986}]{mur86} Murakami, T., Koyama, K., Inoue, H., \& Agrawal, P.~C.\ 1986, \apjl, 310, L31
\bibitem[\protect\citeauthoryear{Naz{\'e} et al.}{2011}]{naz11} Naz{\'e}, Y., Broos, P.~S., Oskinova, L., et al.\ 2011, \apjs, 194, 7 
\bibitem[\protect\citeauthoryear{Naz{\'e} et al.}{2014}]{naz14} Naz{\'e}, Y., Petit, V., Rinbrand, M., et al.\ 2014, \apjs, 215, 10 (+ erratum \apjs, 224, 13)
\bibitem[\protect\citeauthoryear{Naz{\'e} et al.}{2017}]{naz17} Naz{\'e}, Y., Rauw, G., \& Cazorla, C.\ 2017, A\&A, 602, L5 
\bibitem[\protect\citeauthoryear{Naz{\'e} \& Motch}{2018}]{naz18} Naz{\'e}, Y., \& Motch, C.\ 2018, \aap, 619, A148
\bibitem[\protect\citeauthoryear{Nebot G{\'o}mez-Mor{\'a}n et al.}{2013}]{neb13} Nebot G{\'o}mez-Mor{\'a}n, A., Motch, C., Barcons, X., et al.\ 2013, \aap, 553, A12 
\bibitem[\protect\citeauthoryear{Nebot G{\'o}mez-Mor{\'a}n et al.}{2015}]{neb15} Nebot G{\'o}mez-Mor{\'a}n, A., Motch, C., Pineau, F.-X., et al.\ 2015, \mnras, 452, 884 
\bibitem[\protect\citeauthoryear{Neiner et al.}{2011}]{nei11} Neiner, C., de Batz, B., Cochard, F., et al.\ 2011, \aj, 142, 149 
\bibitem[\protect\citeauthoryear{Nieva}{2013}]{nie13} Nieva, M.-F.\ 2013, \aap, 550, A26 
\bibitem[\protect\citeauthoryear{Postnov et al.}{2017}]{pos17} Postnov, K., Oskinova, L., \& Torrej{\'o}n, J.~M.\ 2017, \mnras, 465, L119
\bibitem[\protect\citeauthoryear{Rauw et al.}{2013}]{rau13} Rauw, G., Naz{\'e}, Y., Spano, M., Morel, T., \& ud-Doula, A.\ 2013, \aap, 555, L9
\bibitem[\protect\citeauthoryear{Rauw et al.}{2018}]{rau18} Rauw, G., Naz{\'e}, Y., Smith, M.~A., et al.\ 2018, \aap, 615, A44
\bibitem[\protect\citeauthoryear{Reig}{2011}]{rei11} Reig, P.\ 2011, \apss, 332, 1
\bibitem[\protect\citeauthoryear{Rivinius et al.}{2003}]{riv03} Rivinius, T., Baade, D., \& {\v{S}}tefl, S.\ 2003, \aap, 411, 229
\bibitem[\protect\citeauthoryear{Robinson et al.}{2002}]{rob02} Robinson, R.~D., Smith, M.~A., \& Henry, G.~W.\ 2002, \apj, 575, 435
\bibitem[\protect\citeauthoryear{Sana et al.}{2006}]{san06} Sana, H., Rauw, G., Naz{\'e}, Y., et al.\ 2006, \mnras, 372, 661
\bibitem[\protect\citeauthoryear{Smith et al.}{2016}]{smi16} Smith, M.~A., Lopes de Oliveira, R., \& Motch, C.\ 2016, Advances in Space Research, 58, 782
\bibitem[\protect\citeauthoryear{Smith et al.}{2017}]{smi17} Smith, M.~A., Lopes de Oliveira, R., \& Motch, C.\ 2017, \mnras, 469, 1502
\bibitem[\protect\citeauthoryear{Tsujimoto et al.}{2018}]{tsu18} Tsujimoto, M., Morihana, K., Hayashi, T., et al.\ 2018, \pasj, 70, 109
\bibitem[\protect\citeauthoryear{ud-Doula et al.}{2014}]{udd14} ud-Doula, A., Owocki, S., Townsend, R., et al.\ 2014, \mnras, 441, 3600
\bibitem[\protect\citeauthoryear{ud-Doula et al.}{2018}]{udd18} ud-Doula, A., Owocki, S.~P., \& Kee, N.~D.\ 2018, \mnras, 478, 3049
\bibitem[\protect\citeauthoryear{Wegner}{2007}]{weg07} Wegner, W.\ 2007, \mnras, 374, 1549 
\end{thebibliography}
\end{document}